\documentstyle[preprint,prb,aps]{revtex}

\newcommand{\be}{\begin{equation}}
\newcommand{\ee}{\end{equation}}
\newcommand{\bea}{\begin{equationarray}}
\newcommand{\eea}{\end{equationarray}}
\newcommand{\ba}{\begin{array}}
\newcommand{\ea}{\end{array}}

\begin{document}

\title{ Vortex Dynamics within the BCS Theory   }

\author{P. Ao }
\address{ Department of Theoretical Physics,
              Ume\aa{\ }University, S-901 87, Ume\aa, SWEDEN }
\author{X.-M. Zhu }
\address{ Department of Experimental Physics,
          Ume\aa{\ }University, S-901 87, Ume\aa, SWEDEN   }

\maketitle


\begin{abstract}
Based on the BCS theory 
we outline a conventional path integral derivation of the transverse force 
and the friction for a vortex in a superconductor.
The derivation is valid in both clean and dirty limits 
at both zero and finite temperatures.
The transverse force is found to be precisely 
as what has been obtained 
by Ao and Thouless using the Berry's phase method. 
The friction is essentially the same as the Bardeen and Stephen's result.
\footnote{ Supported by Swedish NFR. To appear in Physica C.}
\end{abstract}

\maketitle

\noindent
{\bf 1. Introduction } 

In condensed matter there are two types of important excitations:
the gapless and elementary ones, such as spin waves or phonons,
which determine the local properties, 
and the topological ones, domain walls or dislocations, 
which determine the global properties.
Vortices in superconductors belong to the second category.
They have been under intensive studies since earlier sixties\cite{brandt}.
Still, there has no general agreement  
on form of the vortex dynamics 
equations.\cite{brandt,ao1,thouless,bardeen,ao2,kopnin,simanek,feigelman}
We point out here that the main disagreement
among many microscopic derivations 
lies in the use or not of the relaxation 
time approximation to account for the effects of impurities
in the force-force correlation function. 
To provide a natural treatment of the both the transverse force 
and the friction 
and to avoid the use of this approximation, 
we present a  path integral derivation within the BCS theory.
The important results here are  
that the transverse force is the same as obtained by 
the Berry's phase method\cite{ao1} and by the total
force-force correlation function method\cite{thouless}, 
and  is indeed insensitive to 
impurities, and that the friction coincides essentially with the Bardeen and 
Stephen result\cite{bardeen} but differs in some important details. 
Specifically, the nature of the friction is determined by the
spectral function of the Hamiltonian. 
The relevance of the present theory to 
experiments has been discussed elsewhere.\cite{zhu} 

\noindent
{\bf  2. Path Integral Formulation }

We begin with the standard BCS Lagrangian in the imaginary time representation
for s-wave pairing. We will only consider neutral superconductors here,
but the coupling to electromagnetic fields does no affect 
our results in the extreme type II case. The Lagrangian is
\be
\begin{array}{ll}
L_{BCS}  = &  \left. \sum_{\sigma}\psi^{\dag}_{\sigma}( x,\tau) 
      \right(  \hbar\partial_{\tau}- \mu_F  \\
      & - \left.
      \frac{\hbar^{2}}{2m} \nabla^{2} + V(x)  \right) \psi_{\sigma}(x,\tau) \\
      & - g\psi^{\dag}_{\uparrow}(x,\tau) \psi^{\dag}_{\downarrow}(x,\tau)
          \psi_{\downarrow}(x,\tau) \psi_{\uparrow}(x,\tau ) \; ,
\end{array} 
\ee
where $\psi_{\sigma}$ describes electrons with spin $\sigma=(
\uparrow, \downarrow )$, $\mu_F$ the chemical potential determined by the 
electron number density, $V(x) $ the impurity potential, and $x =(x,y,z)$.
A vortex at $x_v$ has been implicitly assumed. 
The partition function is 
\be
\ba{cl}
   Z = & \int {\cal D}\{x_v, \psi^{\dag}, \psi \} \times             \\
       &    \exp\left\{ - \frac{1}{\hbar} \int_0^{\hbar\beta } 
             d\tau \int d^3x L_{BCS} \right\}   \; ,
\ea      
\ee
with $\beta = 1/k_B T $ , and $d^3x = dxdydz$.
Inserting the identity in the functional space,
\[ 
   \begin{array}{lc}
   1 = & \left. \int {\cal D}\{ \Delta^{\ast}, \Delta \}
       \exp \right\{- \frac{g}{\hbar} \int_0^{\hbar\beta } 
      d\tau \int d^3x\times\\
       & \left| \psi_{\downarrow} \psi_{\uparrow}
     + \left.  
         \frac{1}{g} \Delta (x,\tau) \right|^2  \right\} \; ,
   \end{array}
\]
into Eq.(2) we have 
\[
  \begin{array}{cl}
   Z   =  & \int {\cal D}\{x_v, \psi^{\dag}, \psi, 
        \Delta^{\ast}, \Delta \} \times   \\   
     &  \exp \left\{ 
        - \frac{1}{\hbar} \int_0^{\hbar\beta } d\tau \int d^3x
           \right. \times  \\
     &  \left( \psi^{\dag}_{\uparrow}, 
              \psi_{\downarrow} \right )
       ( \hbar\partial \tau + {\cal H} )  \left( \begin{array}{c} 
                       \psi_{\uparrow}               \\
                       \psi^{\dag}_{\downarrow}
                       \end{array}  \right)   \\
      & \left. - \frac{1}{g}  \int_0^{\hbar\beta } d\tau \int d^3x  
         |\Delta |^2   \right\} \; . 
   \end{array}
\]
Here the Hamiltonian is defined as
\be
   \ba{l}
   {\cal H}( \Delta, \Delta^{\ast} ) = \left( \begin{array}{cc} 
                     H & \Delta  \\
                    \Delta^{\ast} & - H^{\ast} 
                   \end{array} \right) \; ,
   \ea
\ee
with $H =  - (\hbar^{2}/2m ) \nabla^{2} -  \mu_F + V(x)$.
Integrating out the electron
fields $\psi_{\sigma}^{\dag}$ and $ \psi_{\sigma}$ first, then integrated out 
the the auxiliary(pair) fields $\Delta$ under the meanfield approximation, 
one obtains the partition function for the vortex 
\be
   \ba{l}   
   Z = \int {\cal D}\{ x_v \}
       \exp \left\{ - \frac{ S_{eff} }{ \hbar }  \right\} \; ,
   \ea
\ee 
with the effective action
\be
   \ba{l}
   \frac{S_{eff} }{\hbar} = - Tr \ln G^{-1} -
     \frac{1}{\hbar g}\int_0^{\hbar\beta} d\tau \int d^{3}x|\Delta|^{2} \; ,
   \ea 
\ee
where $Tr$ includes internal and  space-time indices,
and the Nambu-Gor'kov (NG) Green's function $G$ defined by 
\be
    \ba{l}
     ( \hbar \partial_\tau + {\cal H})
      G(x,\tau; x',\tau') = \delta(\tau-\tau') \delta^3(x-x') , 
    \ea
\ee
together with the BCS gap equation, or the self-consistent equation,
\be
  \ba{l}
     \Delta(x,\tau) = - g \; <\psi_{\downarrow}(x,\tau)
                              \psi_{\uparrow}  (x,\tau) >    \; .
   \ea
\ee
A special attention should be paid to the equal time limit of the 
NG Green's function.\cite{schrieffer}

We assume that the vortex is confined to move in a small regime around a point 
at $x_0$, 
which allows a small parameter expansion in terms of the difference 
between the vortex position $x_v$ and $x_0$. 
We look for the long time behavior of vortex dynamics under this small
parameter expansion. 
For the meanfield value of the order parameter, this expansion is
\be
   \ba{ll}
   \Delta(x,\tau, x_v) = & \left( 1 
         + \delta x_v(\tau) \cdot \nabla_0 \right. \\
    & \left. + \frac{1}{2}  ( \delta x_v(\tau) \cdot \nabla_0 )^2 \right) 
         \Delta_0(x,x_0) \; .
   \ea
\ee
Here $\delta x_v = x_v - x_0$.
In Eq.(8) we have used the fact that when $x_v = x_0$ 
the vortex is static.
The effective action for the vortex to the same order is, 
after dropping a constant term,
\be
   \ba{lc}
     \frac{S_{ eff } }{\hbar }  = & - \frac{1}{2} Tr (G_0 \Sigma' )^2 
           + \frac{1}{\hbar g} \int_0^{\hbar\beta} \tau \int d^3x \times \\
       &   \delta x_v\cdot \nabla_0 \Delta^{\ast}_0 \; 
                 \delta x_v\cdot \nabla_0 \Delta_0 \; ,
   \ea 
\ee
with
\be
   \ba{l} 
    \Sigma' = \left( \begin{array}{cc}
                    0 &  \delta x_v\cdot \nabla_0 \Delta_0 \\
             \delta x_v\cdot \nabla_0 \Delta^{\ast}_0 & 0 
                    \end{array} \right)  \; .
   \ea
\ee
Here $G_0$ is the NG Green's function with $\Delta(\Delta^{\ast})$
replaced by  $\Delta_0(\Delta^{\ast}_0)$,
and the gradient $\nabla_0$ is with respect to $x_0$.

Now we construct the NG Green's function $G_0$ 
following the usual procedure. 
First, we consider the eigenfunctions of the Hamiltonian 
${\cal H}_0 = {\cal H}(\Delta_0, \Delta^{*}_0) $. 
The stationary Schr\"{o}dinger equation, 
the Bogoliubov-de Gennes equation, is 
\be
   \ba{l}
   {\cal H}_0 \Psi_k(x) = E_k \Psi_k(x) \; ,
   \ea
\ee
with 
\[
   \ba{l}
   \Psi_k(x) = \left( \begin{array}{c} u_k(x) \\ v_k(x) \end{array} 
       \right) \; .
   \ea
\]   
Since ${\cal H}_0$ is hermitian, all its eigenstates form a  
complete and orthonormal set, that is, 
\[ 
   \ba{l}
   \int d^3 x \Psi^{\dag}_k(x)\Psi_{k'}(x)  
    = \delta_{k,k'} \; ,
   \ea
\]
and 
\[
  \ba{l}
  \sum_k \Psi_k(x)\Psi_k^{\dag}(x')           
     =  \left( \begin{array}{cc}
         1 & 0 \\
         0 & 1 
       \end{array} \right) \delta^3(x-x')  \; , 
  \ea
\]
with $\Psi^{\dag}(x) = ( u^{\ast}(x), v^{\ast}(x) ) $.
Furthermore, Eq.(11) has an interesting property: 
\[
  \ba{l}
    {\cal H}_0 \Psi(x) = E \; \Psi(x)     \Rightarrow 
    {\cal H}_0 \overline{\Psi}(x) = - E \; \overline{\Psi}(x) \; ,
  \ea
\]
with 
\[
  \ba{l}
  \overline{\Psi}(x) = \left( \begin{array}{c} 
                                v^{\ast}(x) \\ 
                      - u^{\ast}(x)\end{array} \right) \; .
  \ea
\]
Finally, since $\delta x_v \cdot \nabla_0 {\cal H}_0 = \Sigma'$, 
for $k\neq k'$ we have 
\[
   \ba{l}
    \int d^3x \Psi_k^{\dag}(x)  \Sigma' \Psi_{k'}(x) \\
      =  (E_{k'} - E_{k} ) \delta x_v \cdot \int d^3x \Psi_k^{\dag}(x)  
      \nabla_0\Psi_{k'}(x)  \; ,
   \ea  
\]
which implies that the two ways of specifying the vortex 
coordinate, through the trapping potential or through the order parameter,
are equivalent. 
Those relationships will be used below.

Given the eigenfunctions of Eq.(11), the NG Green's 
$G_0$ can be expressed as
\[
  \ba{l}
   G_0(x,\tau; x',\tau' ) = \sum_{n,k} \frac{-1}{\hbar \beta} 
          \frac{ e^{ - i \omega_n (\tau -\tau') } }{ i\hbar \omega_n - E_k }
          \Psi_k(x) \Psi_k^{\dag}(x') .
  \ea
\]
Here $\omega_n = n \pi / \hbar\beta$, with $n$ odd integers.
Assuming the rotational symmetry after the impurity average,
a straightforward calculation leads to the following effective action 
\be
   \begin{array}{l}  
    S_{eff} = \left. \frac{1}{2}\int_0^{\hbar\beta} d\tau 
     \right[ K \; |\delta x_v(\tau) |^2 \\
               + \int_0^{\hbar\beta} d\tau' 
      F_{\parallel} ( \tau-\tau')|\delta x_v(\tau)- \delta x_v(\tau') |^2  \\
               + \left.  \int_0^{\hbar\beta} d\tau' 
            F_{\perp} ( \tau-\tau') 
           (\delta x_v(\tau) \times \delta x_v(\tau') )\cdot \hat{z} 
                  \right]  \; ,
    \end{array}
\ee
with the spring constant in the effective potential,
\be
   \ba{l}
    K =  \frac{1}{g} \int d^3x |\nabla_0 \Delta_0^{\ast}(x,x_0)|^2
        - \int^{\infty}_{0} d\omega \frac{ J(\omega) }{\omega} \; ,
   \ea 
\ee
the longitudinal correlation function,
\be
   \ba{l} 
    F_{\parallel}(\tau) = \frac{1}{\pi} \int^{\infty}_{0} d\omega J(\omega)
     \frac{ \cosh[\omega(\frac{\hbar\beta }{2} - |\tau|)] }  
          { \sinh [\omega \frac{\hbar\beta }{2}]          }  \; ,
   \ea 
\ee
and the transverse correlation function, in the long time limit,  
in terms of the virtual transitions,
\be
   \ba{ll}
   F_{\perp}(\tau) = & - \partial_{\tau-\tau'} \delta(\tau-\tau') 
                       \sum_{k,k'} \int d^3x \int d^3x'  \\
    & \hbar ( f_k - f_{k'} )  \frac{1}{2} \hat{z}\cdot
       \left(  \Psi_k^{\dag}(x') \nabla_0\Psi_{k'}(x')
        \times   \right.    \\
    &  \left. \nabla_0\Psi_{k'}^{\dag}(x)  \Psi_k(x) \right) \; ,
   \ea 
\ee
or in terms of the contribution from each state,
\be
   \ba{ll}
   F_{\perp}(\tau) = & - \partial_{\tau-\tau'} \delta(\tau-\tau') \sum_{k}
                                    \int d^3x \; \hat{z}\cdot \\
     & \hbar ( f_k \nabla_0u_k^{\ast}(x)\times \nabla_0 u_k(x)   \\
     & -(1-f_k) \nabla_0v_k^{\ast}(x)\times \nabla_0 v_k(x) )  \; .\\
    \ea
\ee
To reach Eqs.(12-16), following two identities have also been used:
\[
  \ba{l}
     \sum_n \frac{ e^{ -i \omega_n \delta}  }
                 { i\hbar \omega_n - E_k    } = 
     \left\{
        \ba{cc}
       \beta \; f_k \; ,      & \delta = 0^-  \\
       - \beta \; (1 - f_k )\; , & \delta = 0^+  \\    
         \ea
   \right. \; ,
  \ea
\]
and
\[
  \ba{l}
         \sum_{n-n'} \frac{ - 1}{\beta} 
      \frac{ e^{ - i (\omega_n - \omega_{n'}) \tau} }
           { i \hbar(\omega_n - \omega_{n'} ) - E   } 
        = \frac{1}{2} 
        \frac{ \cosh[ \frac{E }{\hbar} ( \frac{\hbar\beta}{2} - |\tau| ) ] }
             { \sinh[ E \frac{\beta}{2} ]             }  \;,
  \ea
\]    
with the Fermi distribution function $f_k = 1/(1 + e^{\beta E_k} )$, and the 
spectral function
\be
   \ba{lc}
   J(\omega) = & \frac{\pi }{4} \sum_{k,k'} 
     \delta(\hbar\omega - | E_k - E_{k'} |)
     |f_k - f_{k'}|\times \\
   & |\int d^3x \Psi_k^{\dag}(x)\nabla_0{\cal H}_0 \Psi_{k'}(x)|^2 \; .
   \ea  
\ee
  
In the following we discuss the implications of 
$K$, $ F_{\parallel}$, and $ F_{\perp}$ one by one, and show
that Eq.(12) contains both the dissipative effect corresponding to 
the Bardeen-Stephen result\cite{bardeen} and the transverse force
identical to the one by Berry's phase method\cite{ao1} or the total
force correlation function method\cite{thouless}.

\noindent
{\bf 3. Correlation Functions } 

\noindent
{\it 3.1 Effective Spring Constant K }

For the purpose of getting the friction and the transverse force,
the precise value of $K$ is irrelavent.

\noindent
{\it 3.2 Longitudinal Correlation Function $F_{\parallel}$ } 

This longitudinal correlation function 
contains all information on the vortex friction, revealed by the fact that 
Eq's.(12,14) are identical to  the time-nonlocal action  in the quantum 
dissipative dynamics\cite{leggett}.
The friction is determined by the 
low frequency behavior of the spectral function $J(\omega)$, Eq.(17).
The temperature is naturally incorporated into the spectral function.
This form of the action has been used in the study of the 
vortex tunneling.\cite{at}
  
If we classify the eigenstates of the 
Bogoliubov-de Gennes equation, Eq.(11),
according to core(localized) and extended states,
the dissipation comes from all three parts: 
core-to-core, core-to-extended (or extended-to-core), 
and extended-to-extended contributions.
Those contributions are added up, because they are all positive, which
differs from the situation for the transverse correlation function.

For a clean and neutral superconductor, the core states are separated by finite
energy gaps from each other. 
Their contributions to the dissipation will be exponentially small in the slow
motion case.
The dominant contribution in this case comes from the extended states,
because their energy spectrum is continuous. 
This is also true for a thin film after the consideration of
the electro-magnetic field effects.
Those contributions correspond to super-Ohmic cases.\cite{niu}

The presence of impurities mixes up all the clean limit eigenstates, 
and generates a continuous distribution of the energy, even for the
core states, after the impurity average.
A perturbative calculation shows that it is indeed the case,
and the spectral function becomes Ohmic.\cite{az}
This suggests that in the clean limit the dissipation is super-Ohmic, 
and turns into Ohmic in the dirty limit.
Hence in the dirty limit one should expect the 
Bardeen-Stephen result\cite{bardeen}, 
because they have treated the core as in a normal 
state, corresponding to the uniform distribution of the energy eigenvalue.

\noindent
{\it 3.3 Transverse Correlation Function $F_{\perp}$ }

In the long time limit the
there are two equivalent forms for the correlation function:
The virtual transition expression, Eq.(15), and 
the individual state contribution, Eq.(16). 
Their equivalence has already been discussed in 
Ref.\cite{ao2} with the aid of conservation laws.
Eq.(15) can be again casted into various equivalent forms 
because of the cancelations among those virtual transitions.
For example, it can be expressed only in terms of the core-to-core 
transitions. 
In this case it is a statement of the spectral flow\cite{ao2}.
But one can also cast it into the forms of core-to-extended, 
or extended-to-extended transitions.
 
In Eq.(16), the counting of individual state contributions is 
expressed as an area integral of the momentum commutator. 
If we express it as the line integral of the momentum density 
far away from the core,
the immediate conclusion, as drawn in Ref.\cite{thouless}, is that
localized core states do not contribute, because the momentum density
at the core is zero. 
The insensitive to the impurities is most transparent from Eq.(16):
In the one-body density matrix both the electron number 
density and the phase $\theta(x - x_v)$  
are all insensitive to impurities.
Here the phase $\theta$ is defined through the order parameter 
$\Delta(x,t, x_v) \rightarrow  |\Delta| e^{ i q \theta(x-x_v) } $, with 
$q = \pm 1 $ describing the vorticity along the $\hat{z}$ direction 
and $\theta(x) = \arctan(y/x)$.  
Hence, as long as the localization effects due to impurities are negligible,
such as in the cases of usual dirty superconductors, 
the transverse force will not be influenced by impurities.
An analogy can be drawn to electrons in semiconductors and 
in metallic rings: The
momentum is insensitive to band structures and disorder potentials.

\noindent
{\bf 4. Discussions}

 1. The major difference between the present approach and those 
in Ref's.\cite{kopnin,simanek} lies in the use or not of
the relaxation time approximation in the
calculation of the force-force correlation function.
It is found in Ref's.\cite{kopnin,simanek} that this approximation is needed 
to makes the friction finite, which at the same time 
reduces the magnitude of the transverse force.
Early studies of this approximation
have shown that it is in general ambiguous and problematic
in the force-force correlation function\cite{kubo}.
The source for the ambiguity is that it is usually unable
to know {\it a priori} the transverse, the friction, 
and the fluctuating forces, although the total force is often well defined. 
This is exactly what happens in the present situation.

In the present approach without this approximation
we are able to obtain the friction determined by the spectral function 
of the Hamiltonian, and demonstrate that   
the transverse force is insensitive to impurities.

2. 
The explicit counting of each state contribution, Eq.(16), 
confirms that the contribute to the transverse force comes from a region 
away from the core. 
One way to show it, as discussed above, is from the consideration
of the momentum density, 
which is zero at the phase singular point, the vortex position.
In an equivalent way, if the one-density density matrix,
defined by $\{u_k, v_k^{\ast}\}$, is finite at the vortex position, as
allowed by the Bogoliubov-de Gennes equation, the corresponding 
$u_k$ or $v_k^{\ast}$ will not have the phase factor $\theta$.\cite{bardeen2}  
Hence, there is no an additional contribution to the transverse force from the 
phase singular point as supposed in Ref.\cite{feigelman}.

\end{document}